\documentclass[12pt]{article}

\usepackage{amssymb}
\def\labelmark{}
\def\void{}

{\ifx\void\labelname\def\junk{\end{displaymath}}
\else\def\junk{\end{eqnarray}}\fi\junk\labelmark\def\labelname{}}

\newcommand{\bra}{\begin{array}}
\newcommand{\era}{\end{array}}
\newcommand{\beq}{\begin{equation}}
\newcommand{\eeq}{\end{equation}}
\newcommand{\beqa}{\begin{eqnarray}}
\newcommand{\eeqa}{\end{eqnarray}}

\font\mybb=msbm10  at 12pt
\def\bb#1{\hbox{\mybb#1}}

\font\mybbi=msbm10  at 9pt
\def\bbi#1{\hbox{\mybbi#1}}

\def\BC{\bb C}
\def\_\BC{\bbi C}

\newcommand{\om}{\omega}

\newcommand{\la}{\lambda}
\newcommand{\si}{\sigma}

\newcommand{\eps}{\epsilon}

\newcommand{\ga}{\gamma}
\newcommand{\te}{\theta}

\newcommand{\pa}{\partial}
\newcommand{\al}{\alpha}

\newcommand{\st}{\star}
\newcommand{\ti}{\tilde}
\newcommand{\da}{\dagger}

\newcommand{\ov}{\over}
\newcommand{\hb}{\hbar}

\newcommand{\sq}{\sqrt}
\newcommand{\no}{\noindent}
\newcommand{\ev}{\equiv}
\newcommand{\lb}{\label}
\newcommand{\de}{\delta}

\newcommand{\PR}[1]{ {\it Phys.~Rev.} {\bf #1}}
\newcommand{\PRL}[1]{ {\it Phys.~Rev.~Lett.} {\bf #1}}

\newcommand{\JMP}[1]{ {\it J. Math.~Phys.} {\bf #1}}

\def\lst2{{(l^*)^2}}

\def\eqa{\begin{eqnarray}}
\def\eea{\end{eqnarray}}

\begin{document}
\begin{titlepage}
\setcounter{page}{1}
\renewcommand{\thefootnote}{\fnsymbol{footnote}}

\begin{flushright}
hep-th/0209259
\end{flushright}

\vspace{6mm}
\begin{center}

{\Large\bf Noncommutativity Parameter 
and Composite Fermions}

\vspace{12mm}
{\large\bf Ahmed Jellal
\footnote{E-mail: {\textsf jellal@gursey.gov.tr }}}

\vspace{5mm}
{\em Institut f\"ur Physik, 
Technische Universit\"at Chemnitz,\\
D-09107 Chemnitz, Germany}\\

\end{center}

\vspace{5mm}

\begin{abstract}

We determine some particular values
of the noncommutativity parameter $\te$ 
and show that the Murthy-Shankar
approach is in fact a particular case of a more general one.
Indeed, using the fractional quantum Hall
effect (FQHE) experimental data, we give a measurement of $\te$. This 
measurement can be obtained 
by considering some values of the filling factor $\nu$
and other ingredients, magnetic field $B$
and electron density $\rho$. 
Moreover, it is found that $\te$ can be quantized either
fractionally or integrally in terms of the magnetic length $l_0$
and the quantization is exactly what Murthy and Shankar
formulated recently for the FQHE. On the other hand, we show that 
the mapping of the FQHE in terms of the composite fermion basis
has a noncommutative geometry nature and therefore there
is a more general way than the 
Murthy-Shankar method to do this mapping.

\end{abstract}

\vfill
\begin{flushleft}
\end{flushleft}
\end{titlepage}

\newpage

\section{Introduction} 

Laughlin's wavefunctions
\cite{laughlin1}
\begin{equation}
\lb{lau}
\Psi^{\rm L} = \prod_{i<j} (z_i-z_j)^{m} 
e^{-{1\ov 4}\sum_i |z_i|^2}
\end{equation}
actually are good wavefunctions for describing the
fractional quantum Hall effect (FQHE)~\cite{tsui,prange}
at filling factor $\nu={1\ov m}$,
$m$ odd integer. However, the situation at most other
filling factors is somewhat less clear. Several
attempts are proposed to extend Laughlin's theory
by adopting different approaches and assumptions. 
In particular, Jain~\cite{jain, heinonen} 
introduced the composite fermion (CF)
concepts. Indeed, Jain's idea is to explain the
FQHE in terms of the integer quantum Hall effect (IQHE) 
by using the attached
flux notion where each 
electron is assumed to be surrounded by an integer number of flux.
Subsequently, by constructing a
velocity operator in terms of the standard 
operator momentum and weakened vector potential,
Murthy and Shankar~\cite{shankar1,shankar2} proposed
a Hamiltonian formalism for the FQHE mapped in terms of
the CF degrees of freedom.

Recently with Dayi, we proposed~\cite{jellal1} 
an approach based on noncommutative 
geometry tools~\cite{connes} to describe the FQHE 
of a system of electrons. In fact, 
the corresponding filling factor is found to be 
\beq
\lb{dj}
\nu_{\rm DJ} = {\pi\ov 2} \rho ( l^2 - \te )
\eeq
which is identified with the observed fractional values 
$f=1/3,\ 2/3,\  1/5,\ \cdots$. This
approach also allowed us to make a link with
the CF approach 
\cite{jain,heinonen}
of the FQHE by setting an effective
magnetic field
\beq
B_{\rm DJ} = {B\ov 1-\te l^2}
\eeq
similar to that felt by the CF's.

In this paper we would like to return to
our former work~\cite{jellal1}  in order 
to add some relevant applications. Indeed,
by considering the experimental data of
two different systems exhibiting the FQHE,
we determine explicitly the corresponding values of  
the noncommutativity parameter $\te$. Under some
assumptions, we find that  $\te$ can be quantized
in terms of the magnetic length
and the quantization is nothing but what Murthy and Shankar
defined when dealing with the FQHE in terms of the CF's. 
Moreover, we present a generalization
of the Murthy-Shankar approach for the FQHE.

Section 2 is a review of the derivation of 
the Hall
conductivity of a two dimensional system 
of electrons subject to an external magnetic
field and living on both planes,
commutative and noncommutative.
These serve as a guide in section 3 
in order to determine some particular
values of $\te$ and in the meantime 
quantize it.
In section 4 after recalling briefly
the Murthy-Shankar approach, we show that
this approach has a noncommutative
nature and therefore there is a more
general approach.

\section{Hall conductivity}

In this section we shall review the
determination of the Hall conductivity for
a two dimensional system of electrons
subject to a magnetic field $B$.
In fact, we start by recalling the commutative case and 
end up with the noncommutative one.
 
\subsection{Commutative plane}

A system of an electron living on the plane $(x,y)$ 
and in the presence of an uniform external  
$B$ and ${E}$ fields can be described
by the following Hamiltonian
\beq
\bra{l}
\lb{hami2}
H=
\frac{1}{2m}\left[
\left( p_{x} -\frac{eB}{2c} y \right)^2 +
\left( p_{y} +\frac{eB}{2c} x \right)^2 \right] +eEx\\ 
\era
\eeq
where the gauge is chosen to be symmetric 
${\vec A}={B\over 2}(-y,x)$
and the scalar potential is fixed to be
$\phi =-xE_x$.

$H$ can be diagonalised simply by
considering a couple of creation and annihilation
operators. Then, let us define the first pair~\cite{jellal1}
\beq
\bra{l}
\lb{eop}
b^\da =-2i{p}_{\bar z}+{eB\ov 2c}{z}+\la\\
b =2i {p}_{z}+{eB\ov 2c}{{\bar z}}+\la\\
\era
\eeq
and also the second
\beq
\bra{l}
d=2i {p}_{z}-{eB\ov 2c} {{\bar z}}\\
d^{\da}=-2i {p}_{\bar z}-{eB\ov 2c} {z}
\era
\eeq     
where $\la={mcE\ov B}$ and $z=x+iy$ is the complex coordinate.
These sets satisfy the commutation relations
\beq
\bra{l}
\lb{cr}
[b, b^{\da}]=  2m\hb\om \\

[d^{\da},d]=  2m\hb\om\\
\era
\eeq
where $\om= {eB\over mc}$ is the cyclotron frequency. 
The other commutators vanish.
By using the above operators, we can write ${H}$ 
as
\beq
\bra{l}
\label{hc}
{H}= {1\ov 4m}(b^{\da}b + bb^{\da})-
{\la\ov 2m}(d^{\da}+d)-{\la^2\ov 2m}.
\era
\eeq

From the eigenvalue equation 
\beq
H \Psi = E \Psi
\eeq
we obtain eigenstates and energy spectrum:
\beq
\lb{hfs}
\bra{l}
\Psi_{(n,\al)}\ev 
|n,\al>={1\ov \sqrt{(2m\hb{\om})^{n} n!}}e^{i(\al y+{m{\om}\ov 2\hb}xy)}
({b}^{\da})^{n}|0>
\\
E_{(n,\al)}={\hb\om\ov 2}(2n+1)
- {\hb\la\ov m}\al-{\la^2\ov 2m}
\era
\eeq
where $n=0,1,2\cdots$ and $\al\in  \mathbb{R}$.

The corresponding Hall conductivity 
$\si_{\rm H}$ can be derived by using 
the definition of
the related current operator 
${\vec{J}}$,
such as
\beq
\bra{l}
{{\vec{J}}}=-{e\rho\ov m} ({\vec p }+ {e\ov c}{\vec A})\\
\era
\eeq
where $\rho$ is
the electron density. 
Moreover,
the expectation value of ${{\vec{J}}}$ 
can be calculated with respect to the eigenstates 
$|n,\al>$ (\ref{hfs}). Therefore, 
we obtain
\beq
\lb{ncco}
\bra{l}
< {J}_x >=0 \\
<{J}_y>= - \left({\rho ec\ov B}\right) E.
\era
\eeq
The second equation implies that 
the Hall conductivity $\si_{\rm H}$
is 
\beq
\lb{hc}
\si_{\rm H}= - {\rho ec\ov B}.
\eeq
Using the definition of the filling factor: 
\beq
\nu = 2\pi \rho l_0^2
\eeq
where $l_0={\sq{\hb c\ov eB}}$
is the magnetic length, we can write $ \si_{\rm H}$ as
\beq
\si_{\rm H}= - \nu {e^2\ov h}
\eeq

\subsection{Noncommutative plane}

In this subsection, we review
a generalization~\cite{jellal1} of the last 
section in terms of noncommutative
geometry~\cite{connes}.
Notations will be slightly changed
in order to be coherent with our further analysis.
In doing so,
let us start by 
introducing the noncommutativity between the
spatial coordinates, such as
\beq
[x^{i},x^{j}]=i\te^{ij}
\label{nccoo}
\eeq
where $\te^{ij}=\eps^{ij}\te$ is the 
noncommutativity parameter
and $\eps^{12}=-\eps^{21}=1$. Basically, 
we are forced in this case
to replace $fg(x)=f(x)g(x)$ by the relation
\beq
f(x) \st g(x)=\exp[{i\over 2}\te^{ij}
\pa_{x^{i}}\pa_{y^{j}}]f(x)g(y)|_{x=y}
\label{2}
\eeq
where $f$ and $g$ are two arbitrary functions, supposed to be 
infinitely differentiable.
As a consequence, now we are going to deal with 
quantum mechanics by considering the
following algebra
\beq
\bra{lll}
\lb{deqm}
[x^{i},x^{j}]=i\te^{ij}\\

[p^{i},x^{j}]=-i\de^{ij}\\

[p^{i},p^{j}]=0.
\era
\eeq

Actually, we can write down
the noncommutative version of the Hamiltonian~(\ref{hami2}). 
In doing so, let us notice that $H$
acts on an arbitrary function $\Psi(\vec{r},t)$ as
\beq
\bra{l}
\lb{nhami}
H \st \Psi (\vec{r},t) = H^{\rm nc} \Psi (\vec{r},t)
\era
\eeq
which implies that $H^{\rm nc}$ is
\beq
\bra{l}
\lb{nh}
{H}^{\rm nc} =
\frac{1}{2m}\left[
\left( \ga {p}_{x} -\frac{eB}{2c} {y} \right)^2 +
\left(\ga {p}_{y} +\frac{eB}{2c} {x} \right)^2 \right]
+eE({x}-{\te\ov 2\hb}{p}_{y})\\
\era
\eeq
where $\ga$ is a new parameter and defined to be
$\ga = 1- \te l^{-2}$
and $l=2l_0$.

Now, one can use a similar process as in the previous 
section to diagonalise $H^{\rm nc}$. 
Let us define the following operators
\beq   
\lb{nob}
\bra{l}
{\ti b}^{\da}=-2i \ga {{ p}}_{\bar z}+{eB\ov 2c} {z}+\la_{-} \\
{\ti b} =2i \ga {{ p}}_{z}+{eB\ov 2c} {\bar z}+\la_{-}
\era
\eeq
and 
\beq
\lb{nod}   
\bra{l}
{\ti d}=2i \ga {{p}}_{z}-{eB\ov 2c} {{\bar z}}\\
{\ti d}^{\da}=-2i  \ga {{p}}_{\bar z}-{eB\ov 2c}{z}.
\era
\eeq
The sets of operators $({\ti b},{\ti b}^\da )$ and
$({\ti d},{\ti d}^\da )$ commute with each other. Moreover,
they verify the commutation relations 
\beq   
\bra{l}
\lb{ncr}
[{\ti b} ,{\ti b}^{\da}]=  2m\hb{\ti\om} \\
 
[{\ti d}^{\da},{\ti d}]=  2m\hb{\ti\om}\\
\era
\eeq
where ${\ti\om}$ and the $\la_{\pm}$ 
are given by
\beq
\bra{l}
{\ti\om}=\ga\om \\
\la_{\pm}=\la \pm{em E\te \ov 4\ga\hb}. 
\era
\eeq
To ensure these equations hold and
for further analysis, we assume that
the condition  $\theta\ne l^2$ 
is satisfied. 
In terms of the above creation and annihilation
operators, 
the Hamiltonian ${H}^{\rm nc}$ takes the
form
\beq
\bra{l}
\label{nh1}
{H}^{\rm nc}= {1\ov 4m}({\ti b}^{\da}{\ti b}+
{\ti b}{\ti b}^{\da})
-{\la_{+}\ov 2m}
({\ti d}^{\da}+{\ti d})-{\la_{-}^2\ov 2m}.
\era
\eeq

As before, we can solve the eigenvalue equation
\beq
{H}^{\rm nc} \Psi^{\rm nc} = 
E^{\rm nc} \Psi^{\rm nc}
\eeq
to get the eigenstates: 
\beq   
\lb{nef}  
\Psi_{(n,\al,\te )}^{\rm nc}
\ev|n,\al,\te> =
{1\ov \sqrt{(2m\hb{\ti\om})^{n}n!}}
e^{i(\al y +{m {\ti\om}\ov 2\hb}xy)}
({\ti b}^{\da})^{n}|0>
\eeq
and the corresponding eigenvalues:  
\beq
\bra{l}
E^{\rm nc}_{(n,\al,\te)}=
{\hb{\ti\om}\ov 2}(2n+1)-
{\hb\ga\la_{+}\ov m}\al-{m\ov 2}\la_{-}^2 \\
\era
\eeq
where $n=0,1,2...$ and $\al\in \mathbb{R}$.

The 
conductivity  
resulting from the Hamiltonian 
${H}^{\rm nc}$ is determined by
defining the current operator 
${\vec{J}}^{\rm nc}$ 
on the noncommutative plane as
\beq
\lb{nco}
{{\vec{J}}}^{\rm nc} = -
{e\rho \ga\ov m} ( \gamma {\vec p}+
{e\ov c}{\vec A}+{\vec a})
\eeq
where the ${\vec a}$ vector is
\beq
\bra{l}
{\vec a}=(0,-{m eE\te \ov 2\hb\ga}).
\era
\eeq
Its expectation value is calculated with respect 
to the eigenstates $|n,\al,\te>$
(\ref{nef}) and is found to be
\beq
\lb{ncco}
\bra{l}
<{J}_x^{\rm nc}> =0\\
<{J}_y^{\rm nc}> = -\left(\ga {\rho ec\ov B}\right)E.
\era
\eeq
Therefore, the Hall conductivity on 
the noncommutative plane of electrons,
denoted by $\si_{\rm H}^{\rm nc}$, is
\beq
\lb{nhc}
\si_{\rm H}^{\rm nc}= -\ga {\rho ec\ov B}
\eeq
and as before we can define an effective filling factor
\beq
\nu_{\rm eff}={\ga\Phi_0\rho\ov B} 
\eeq 
corresponding to an effective magnetic field:
\beq
\lb{nmf}
B_{\rm eff}={B\ov \ga}
\eeq
where $\Phi_0={hc\ov e}$ is the unit flux. 
To close this section, let us notice that the  
commutative analysis is recovered if the 
noncommutativity parameter $\te$ is 
switched off.

\section{Measurement and quantization of $\te$ }

Before we start, let us mention that in our work~\cite{jellal1} 
we offered two interpretations for equation~(\ref{nhc}). 
In particular~(\ref{nhc}) can be seen as a result of the FQHE 
at fractional filling factor $f$. Identifying
\beq
\lb{fff}
\nu_{\rm eff}|_{\te=\te_{\rm H}}=f 
\eeq
we find
\beq
\lb{fff1}
{\te_{\rm H}}={2\Phi_0\ov\pi B}\Big(1-f{B\ov\Phi_0\rho}\Big)
\eeq
which tells us when 
$\te$ is fixed to be $\theta_{\rm H}$,
one can envisage the Hall effect on
noncommutative plane as the usual 
fractional quantum Hall effect.

\subsection{Measurement}

Next, we determine explicit values of 
$\te$ by using experimental observations.
Such measurements are possible
since we actually have  
a relation~(\ref{fff1}) governing the present parameters.
Basically, to measure $\te$ one can use experimental data where 
$f$ and the corresponding magnetic field are well-known. 
To do this task,
we should fix the FQHE system and the 
corresponding ingredients. 
For instance, let us consider two different
systems of electrons:\\

{\bf GaAs/AlGaAs heterostructure}:
\\
This system was the subject of many investigations
dealing with the FQHE at low temperature and high mobility. 
In~\cite{boebinger}, the authors 
obtained some measurements by considering
the present system of electron density 
$\rho = 1.5 10^{11} cm^{-2}$. 
Their experimental 
data was reported as follows: The energy gap of the 
FQHE state at $\nu=4/3$ is $0.27 K$ at $B=7.3 T$,
while it is $0.19 K$ at $B=5.9 T$ for  $\nu=5/3$
state.

At this stage, we can have a fixed value of the noncommutativity
parameter. Indeed,
the magnetic length can be measured in terms of the magnetic 
field such as     
\beq
l_0[{\rm m}] = 25.65 10^{-9} [B(T)]^{-1/2}.
\eeq  
On the other hand, 
let us rewrite (\ref{fff1}) as follows 
\beq
\lb{fff2}
{\te_{\rm H}}[{\rm m^2}] =4l_0^2-{2f\ov\pi\rho}.
\eeq
Now, we can have an explicit value of $\te$ corresponding
to the above experimental data. Therefore 
for the $\nu=4/3$ state, we obtain 
\beq
\lb{fff3}
{\te_{\rm H}}|_{\nu=4/3} = 0.2055 \; 10^{-11}{\rm cm^2}  
\eeq
while the $\nu=5/3$ state leads us to have  
\beq
\lb{fff3}
{\te_{\rm H}}|_{\nu=5/3} = 0.2617 \; 10^{-11}{\rm cm^2}.  
\eeq
This is a way to give some hints on spatial noncommutativity.
Moreover, another possibility is given in terms of Aharonov-Bohm
effect, where an experiment is proposed to measure 
$\te$~\cite{jellal1}.\\

{\bf GaAs-Al$_{0.3}$Ga$_{0.7}$As heterostructure}:\\
Here we are going to give a table including 
some experimental results and the corresponding 
measurement of the noncommutativity parameter 
$\te$. The above system is considered in~\cite{chang}
and their results are listed in the table.
Therefore, from the author's observation and
equation~(\ref{fff2}), one can end up with
a table involving a summary of the values
for the quantized Hall conductivity of the
various FQHE 
and the corresponding values of $\te$:

\vskip6mm
\begin{center}
\begin{tabular}{|c|c|c|c|c|c|}
\hline
$\nu$  &  $\rho$ $({\rm cm}^{-2})$ &  
$B$ $ ({\rm kG})$ & $4l_0^2$ $ ({\rm cm}^2)$ &
$\te$ (${\rm cm}^{2}$) & ${\te\ov 4l_0^2}$ \\[2mm] \hline
${1/ 3}$    &  $1.53\; 10^{11}$ & $190$ 
& $0.138510 \; 10^{-11}$ & $0.000257 \; 10^{-11}$ & 0.001855  \\[2mm] \hline
${2/ 3}$    &  $2.42\; 10^{11}$ & $150$ 
& $0.175446 \; 10^{-11}$ & $0.000020 \; 10^{-11}$ & 0.000113 \\[2mm] \hline
${2/ 5}$    &  $2.13\; 10^{11}$ & $220$ 
& $0.119622 \; 10^{-11}$ & $0.000008 \; 10^{-11}$ & 0.000066 \\[2mm] \hline
 ${3/ 5}$    &  $2.13\; 10^{11}$ & $147$ 
& $0.179026 \; 10^{-11}$ & $0.000393 \; 10^{-11}$ & 0.002195 \\[2mm] \hline
 ${5/ 3}$    &  $2.06\; 10^{11}$ & $53 $ 
& $0.496545 \; 10^{-11}$ & $0.018780 \; 10^{-11}$ & 0.037821 \\[2mm] \hline
 ${3/ 7}$    &  $2.13\; 10^{11}$ & $206$ 
& $0.127751 \; 10^{-11}$ & $0.000405 \; 10^{-11}$ & 0.003170\\[2mm] \hline
\end{tabular}
\end{center}
\vskip6mm

\no From this table we observe that the ratio ${\te\ov 4l_0^2}$
is very much smaller than one,
which means that $\te\ll 4l_0^2$.
 Therefore, for this system the corresponding 
effective magnetic field can be approximated as
\beq
B_{\rm eff} \approx B +{B\te\ov 4l_0^2}.
\eeq

\subsection{Quantization}

Once the noncommutativity parameter $\te$ is 
linked to the fractional filling factor~(\ref{fff1}),
then one can ask about the quantization of $\te$
in terms of the magnetic length $l_0$. To clarify
this point, let us demand
that $\si_{H}$ is nothing but referring to the IQHE,
namely
\beq
\si_{H} = -i {e^2\ov h}
\eeq
where $i$ is integer value. Then, (\ref{fff1})
can be written as 
\beq
\lb{mff}
{\te_{\rm H}}= 4 l_0^2 \Big(1-{f\ov i}\Big)
\eeq
this tells us that $\te$ is actually quantized either
fractionally or integrally. Now
we would like to make contact with the Murthy-Shankar
$c^2$ parameter~\cite{shankar1}, which is related to 
the CF theory. Indeed, 
let us consider the case 
where the filling factor $f$ is identified to the 
Jain series 
\beq
\lb{jain}
f = {i\ov 2ip +1}
\eeq
where $p=0,1,\cdots$. Now injecting~(\ref{jain})
in~(\ref{mff}), we find
\beq
\lb{mff2}
{\te_{\rm H}}= 4 l_0^2 \Big({2ip \ov 2ip + 1}\Big).
\eeq
Setting $k=ip=0,1,2,\cdots$, we obtain 
\beq
\lb{mff3}
{\te_{\rm H}}= 4 l_0^2 \Big({2k\ov 2k + 1}\Big).
\eeq
Therefore the fractional value
${2k\ov 2k + 1}$ is exactly the quantity
$c^2$ defined recently by Murthy and Shankar~\cite{shankar1} 
to formulate a Hamiltonian   
for the FQHE in the CF basis. 
Then, we can write the above relation
as 
\beq
\lb{mff4}
{\te_{\rm H}\ov 4 l^2}= c^2.
\eeq
We will come back to the Murthy-Shankar method
in the next section when we will talk about the CF's.

\section{Composite fermions}

In this section, we show that
the recent results obtained by 
Murthy and Shankar concerning the CF's
are particular cases of what is 
derived before~\cite{jellal1} by considering
electrons moving on the noncommutative
plane.
CF's are particles
carrying an even number $2p$ $(p=1,2,\cdots )$ of flux quanta (vortices).
They have the same charge, spin and statistics as the usual particles, but they
differ from them since they experience an effective magnetic field   
\beq
\lb{cfm}
B^*=B \pm 2p\rho\Phi_0.
\eeq

Before going on, we note that
a system of electrons living on  
the noncommutative plane 
in the presence of an external magnetic 
field $B$ can be seen
as a set of CF's subject to an effective
magnetic field $B_{\rm eff}$
and living on the usual plane.
This statement is supported by the following 
relation~\cite{jellal1}  
\beq
\lb{emf}
B_{\rm eff}|_{\te=\te_{\rm c}}=B^{*}
\eeq
where $B_{\rm eff}$ is given in (\ref{nmf}).
This equation leads us to have
\beq
\lb{cte}
\te_{\rm c}={2\Phi_0\ov\pi B}\Big[1\pm(1-2p{\rho\Phi_0\ov B})^{-1}\Big].
\eeq

\subsection{Murthy-Shankar approach}

In this subsection, we are going
to review shortly the recent development
of Murthy and Shankar~\cite{shankar1} for the FQHE.
Indeed, the authors considered a CF Hilbert space,
where 
each fermion is described by a coordinate $\vec r$ and momentum 
$\vec p$, 
by constructing the following operator
\beq
{\vec \pi }={\vec p} +e{\vec A}^*
\eeq
where the weakened vector potential $A^*$
\beq
A^* = {A \over 2ps+1}
\eeq
is what the CF sees, where $p$ and $s$ are integers.
In terms of these variables, the electron guiding 
center $\vec R_e$ takes the form
\beq
{\vec R}_e = {\vec r} -{l^2\over (1+c)}\hat{\vec z}
\times {\vec \pi}
\eeq
where the c parameter is given by
\beq
c^2 = {2ps\over 2ps+1}.
\eeq 
It is easy to see that
\beq
\left[ R_{ex}, R_{ey} \right] = - il_0^2.
\eeq
Actually, $\vec R_e$ can be written in terms of the CF guiding center
and cyclotron coordinates $ \vec R$ and $\vec\eta$, such that 
\beq 
\vec R_e= \vec R+ c\vec\eta.
\eeq
Another pair of guiding center-like coordinates
commuting with $ \vec R_e$ can be defined
\beq 
\vec R_v = \vec R + {1\ov c} \vec \eta 
\eeq
which can also be mapped in terms of
$\vec r$ and ${\vec \pi}$:
\beq
\bra{l}
{\vec R}_v = {\vec r} +{l^2\over c(1+c)}\hat{\vec z}
\times {\vec \pi}
\\ 
\left[ R_{vx}\ , R_{vy} \right] =  il^2/c^2.
\era
\eeq
These correspond to the guiding center coordinates of a particle
of charge $-c^2 = -2ps/(2ps+1)$, which is precisely the charge of an
object that must pair with the electron to form the CF called
{pseudo-vortex} coordinate, since it has the same charge
as a 2s-fold vortex in Laughlin states. Since $\vec R_v$ has a magnetic
algebra charge of $-c^2$, and there is one pseudo-vortex per electron,
one can see that it is always at filling factor:
\beq 
\lb{sm}
\nu'= {\nu\over -c^2} = -{1\over 2s}
\eeq 
corresponding to the bosonic Laughlin wavefunctions~\cite{lee}:
\beq
\Psi^{L}=\prod\limits_{i<j}(z_i-z_j)^{2s} 
e^{-\sum\limits_{i}^{} {c^2|z_i|^2\over4}}
\label{half-bos-laugh}
\eeq

For many speculations about this approach and
related matters, one can see
the author's original work~\cite{shankar1}.

\subsection{Noncommutative nature}

We show that the Murthy-Shankar approach
has a noncommutative nature and therefore
there is a theory more general 
and is actually noncommutativity 
parameter $\te$ dependent. In fact, we have seen 
in the beginning of this section that the
CF theory can be envisaged
as a particular theory of electrons
moving on the NC plane and this statement
is governed by equation~(\ref{cte}).
To process, let us write~(\ref{cte})
as follows
\beq
\lb{cte2}
{\te_{\rm c}\ov 4l_0^2} =            
{4p\pi\rho l_0^2\ov 4p\pi\rho l_0^2 \pm 1}.
\eeq
Remembering that the filling factor is 
given by $\nu= {2\pi\rho l_0^2}$, putting
this in relation~(\ref{cte2}), we find
\beq
\lb{cte3}
{\te_{\rm c}\ov 4l^2} =            
{2p\nu\ov 2p\nu \pm 1}.
\eeq  
In a similar way and for the same reason as
we have seen in the last subsection, let 
us define a noncommutative filling
factor as
\beq
\lb{cff1}
\nu^{'}_{\rm nc} =  {\nu^{*}\ov  -\te_{\rm c}/ 4l^2}
\eeq   
for any filling factor $\nu^{*}$ characterizing
the quantum Hall effect. It is equivalent to
\beq
\lb{cff2}
\nu^{'}_{\rm nc} = - \nu^{*} \Big(1\pm {1\ov 2p\nu}\Big)
\eeq
and the corresponding wavefunctions can be written as
\beq
\Psi^{\rm nc}=\prod\limits_{i<j}(z_i-z_j)^{1/-\nu^{'}_{\rm nc}} \;
e^{-{\te_{\rm c}\ov 16l^2}\sum\limits_{i}^{} {|z_i|^2}}
\label{jel-laugh}.
\eeq
This may be a general way to see that the Murthy-Shankar method 
is in fact a particular case of noncommutative
analysis. To prove this statement, let us
demand that $\nu$ is referred to the IQHE by fixing 
$\nu \ev i= 1,2,\cdots$. Therefore,
we end up with
\beq
\lb{cte4}
{\te_{\rm c}\ov 4l^2}|_{\nu\ev i} =            
{2ip\ov 2ip \pm 1}
\eeq     
showing that
\beq
\lb{cte5}
{\te_{\rm c}\ov 4l^2}|_{\nu^{*}\ev i}  = c^2
\eeq  
Moreover, equation~(\ref{cff2}) becomes 
\beq
\lb{cff3}
\nu^{'}_{\rm nc}|_{\nu\ev i} = - \nu^{*} \Big(1\pm {1\ov 2ip}\Big).
\eeq
Fixing $\nu^{*}$ to be ${i\ov 2ip \pm 1}$, we obtain 
\beq
\lb{cff4}
\nu^{'}_{\rm nc}|_{\nu\ev i} \ev \nu'= -{1\ov 2p}
\eeq
which is nothing but the Murthy-Shankar filling 
factor~(\ref{sm}). Therefore we arrived to conclude
that considering the weakened vector potential $A^*$
seen by CF's is equivalent to having a set of
particles living on noncommutative space. 
Clearly, this analysis gives one example 
among other applications of noncommutative
geometry in physics
and 
shows how NC can serve to study some
condensed matter physics phenomena.

\section{Conclusion} 

By exploring the experimental data
of some fractional quantum Hall systems
a measurement of the noncommutativity parameter $\te$ 
is given. In fact, two different 
heterostructures:  GaAs/AlGaAs and 
GaAs-Al$_{0.3}$Ga$_{0.7}$As are considered and showing different
values of $\te$. For the first system, we
obtained the $\te$ values corresponding to
the filling factors ${4\ov 3}$ and ${5\ov 3}$.
While for the second one
several values are determined and   
a comparison with respect to the magnetic
length was given, see table.  
This measurement gives some
hint on spatial noncommutativity. 

On the other hand, we developed an analysis
in terms of noncommutative geometry 
to generalize the recent Murthy-Shankar proposal
and also to prove that their proposal has in fact
a noncommutative origin.

\section*{Acknowledgments} 

I would like to thank P. Bouwknegt for his kind
invitation to participate in a workshop on
{\it Noncommutative Geometry and Fractional 
Quantum Hall Effect at the University of Adelaide (02-06 August 2002)},
and also for
his warm hospitality during my stay at
his {\it Department of Physics and Mathematical
Physics} where the present work has been completed.
I am grateful to \"O.F. Dayi for his helpful comment 
and P. Dawson for reading the manuscript.

\end{document}